\documentclass[aps,prd,groupedaddress,preprint,amsmath,amssymb,showpacs]{revtex4-1}  

\begin{document}


\title{Neutrino masses in a 331 model with right-handed neutrinos without doubly charged Higgs via
           inverse and double seesaw mechanisms}


\author{E. Cata\~no M.}
\email{ecatanom@unal.edu.co}

\author{R Mart\'inez}
\email{remartinezm@unal.edu.co}
\author{F. Ochoa}
\affiliation{Departamento de F\'isica, Universidad Nacional de Colombia,  Bogot\'a}

\date{\today}

\begin{abstract}
We discuss a 331 model with three scalar triplets and neutral fermion singlets. We show that in the 331 model with right-handed neutrinos, it is possible to obtain small active neutrino masses via the double and inverse seesaw mechanisms, without the use of scalar sextets or triplets with doubly-charged Higgs. Two types of models are discussed. If we have a large Majorana mass matrix for the singlets, the spectrum of neutrinos presents light, heavy and very heavy masses. The other possibility is a small (zero) Majorana mass matrix, which leads to pseudo-Dirac (Dirac) heavy neutrinos in the TeV scale, in addition to the active light neutrinos.

\end{abstract}


\maketitle

\section{Introduction}

Neutrinos have played an important role in the evolution of particle physics, and nowadays the explanation of some of their properties is one of the main goals of models beyond the Standard Model (SM). In the SM, there are three flavours of massless neutrinos, which take part in the charged and neutral current weak interaction, and the states that describe different flavour neutrinos are left-handed (LH), orthogonal, and form a $SU(2)_L$ doublet with the corresponding charged lepton field \cite{Nakamura:2010zzi}.  The existence of three different flavours is a well-established experimental fact, and from existing data it is concluded that they are always produced in weak interaction processes in a state that is predominantly left-handed. We do not have compelling evidence for the existence of predominantly right-handed neutrinos, and therefore conclude that they should be \textit{sterile}, i.e. that their interaction with matter should be much weaker than the weak interaction of LH neutrinos. Formally, this could be expressed by describing the RH neutrino as a $SU(2)_L$ singlet. In extensions of the SM, RH neutrinos serve in the explanation of masses and mixing, as well as the disparity between neutrino and charged lepton and quarks mass scales, and the generation of matter-antimatter asymmetry.

Experiments with solar, atmospheric and reactor neutrinos provide evidence of neutrino oscillations, caused by non-zero mass and mixing matrices \cite{Cleveland:1998nv,Fukuda:1996sz,Abdurashitov:2009tn,Anselmann:1992um,Fukuda:2002pe,Ahmad:2001an,Fukuda:1998mi,Ashie:2004mr,Eguchi:2002dm,Araki:2004mb,Arpesella:2008mt,Ahn:2006zza,Michael:2006rx}. 
Formally, this means that the LH flavour neutrino fields $\nu _{lL} (x), l=e,\mu, \tau$ which enter the expression for the lepton current in the CC weak interaction Lagrangian (\textit{weak eigenstates}), are linear combinations of three or more fields $\nu_j$ with masses $m_l \not=0$ (\textit{mass eigenstates}),
\begin{equation}
\nu_{lL}(x)=\sum _j U_{lj}\nu_{jL} (x),
\end{equation}
where $U$ is a unitary matrix (Pontecorvo-Maki-Nakagawa-Sakata PMNS matrix).
Almost all existing data can be described with at least three light neutrinos, with $m_{1,2,3}\lesssim1eV$ and $m_1 \not= m_2 \not= m_3$, but the existence of additional, sterile neutrinos is not ruled out.

In order to give an explanation to the smallness of neutrino masses, basically two types of models have been attempted: seesaw and radiative corrections \cite{Cortez:2005cp}.
The seesaw mechanism is often denoted as the most elegant scheme  \cite{Dias:2011sq}, and relies in the violation of lepton number at a very high energy scale ($M$), giving a mass with the form 
$m_\nu=\frac{v_w^2}{M}$. There are three basic ways of realizing this mechanism: with a heavy right-handed Majorana neutrino coupled to $\nu_L$ via the SM scalar doublet (canonical, or type I seesaw); with a heavy scalar triplet (type II), or a heavy fermionic triplet (type III). Since the mass scale $M$ associated with the new fields is very high ($\sim 10^{12}$ GeV), these models have the huge disadvantage of not being accessible to experiments.

In the last few years, there have been numerous attempts of accommodating the seesaw mechanism in models in the TeV scale, which could be probed in LHC. Some of these models are based on the $SU (3)_C \otimes SU (3)_L \otimes U(1)_X$ gauge symmetry, and are called 331 models for short. Since the SM 321 group is a subgroup of 331, it is possible to account for the known phenomenology, and try to fit new effects in the windows for new physics. One of the motivations of these models is the explanation of the origin of generations, and the prediction of charge quantization for three-family models. However, even after imposing restrictions, specially the cancellation of anomalies, there remains a free parameter $\beta$  and therefore it is not possible to identify a unique version of a 
331 model. 

 Neutrino masses have been treated before in 331 models as radiative corrections, or including a scalar sextet which leads to Majorana mass terms \cite{Dong:2008sw}. Many of these, particularly those with a scalar sextet, have the characteristic of including doubly-charged Higgs bosons, which may or may not be  observed experimentally. Working with a sextet has a couple of down-sides, namely the huge amount of extra parameters, and the inclusion of a $SU(2)$ triplet with a very large mass.  
 In this work, we show an alternative for generating neutrino mass matrices in a 331 model with $\beta=-\frac{1}{\sqrt{3}}$, which does not involve any exotic charges, neither in the fermionic nor in the scalar sector. With the addition of 331-singlet neutrinos, one can find light neutrino masses within the sub-eV range via the so-called inverse, double and linear seesaw mechanisms \cite{Ibarra:2010xw, Gu:2010xc}.
 
 The article is organized as follows: In section \ref{elements}, we review the basics of $331$ models, and mention previous works that address neutrino masses  with this symmetry. In section \ref{model}, we focus on our particular version, presenting both the scalar and fermionic sector, and the Lagrangian that leads to the three different scenarios for neutrino masses. In appendix \ref{diag} we show the procedure for the diagonalization of the neutrino mass matrix.


\section{\label{elements}Elements of 331 models}

In order to construct a gauge model based on the group $SU(3)_C \otimes SU(3)_L \otimes U(1)_X$,
first of all we need to ensure that it describes the electromagnetic interaction, which is done 
demanding that the sector $SU(3)_L \otimes U(1)_X$ contains the subgroup $U(1)_Q$. The generator $Q$
is defined as
\begin{equation}
 Q=T_3+\beta T_8 +X I,
\end{equation}
where $T_3$ and $T_8$ are diagonal generators of $SU(3)_L$, $X$ is the quantum number associated to $ U(1)_X$ 
and is related to the free parameter $\beta$ \cite{Diaz:2003dk, Diaz:2004fs}. However, it is not possible to identify a unique version of a 
331 model. Some conditions that we have to impose on the model, and that will restrict the possible values for $\beta$
are: that it contains at least the phenomenological particles (with all their degrees of freedom) and the interactions among them, that the anomalies cancel (so that it is a renormalizable theory), that the scalar sector must allow a SSB according to the scheme $331 \rightarrow 321 \rightarrow 31$, and that the extra particles are heavier than the SM ones. One way of choosing $\beta$ is defining the electric charge of the exotic (beyond the SM) particles.

Quarks ($q$) and leptons ($l$) can be in the following representations:
\begin{eqnarray} 
 &\hat{\psi}_L = \begin{cases} 
  \hat{q}_L: (\boldsymbol{3},\boldsymbol{3},X_q^L)= (\boldsymbol{3},\boldsymbol{2},X_q^L) \oplus (\boldsymbol{3},\boldsymbol{1},X_q^L),
  \\
  \hat{l}_L: (\boldsymbol{1},\boldsymbol{3},X_l^L)= (\boldsymbol{1},\boldsymbol{2},X_l^L) \oplus (\boldsymbol{1},\boldsymbol{1},X_l^L),
\end{cases}
\\
 &\hat{\psi}_L^* = \begin{cases} 
  \hat{q}_L^*: (\boldsymbol{3},\boldsymbol{3^*},X_q^L)= (\boldsymbol{3},\boldsymbol{2^*},X_q^L) \oplus (\boldsymbol{3},\boldsymbol{1},X_q^L),
  \\
  \hat{l}_L^*: (\boldsymbol{1},\boldsymbol{3^*},X_l^L)= (\boldsymbol{1},\boldsymbol{2^*},X_l^L) \oplus (\boldsymbol{1},\boldsymbol{1},X_l^L),
\end{cases}
\\
 &\hat{\psi}_R = \begin{cases} 
  \hat{q}_R: (\boldsymbol{3},\boldsymbol{1},X_q^R),
  \\
  \hat{l}_R: (\boldsymbol{1},\boldsymbol{1},X_l^R),
\end{cases}
\end{eqnarray}
where the notation $ (\boldsymbol{3},\boldsymbol{2},X) $, $ (\boldsymbol{1},\boldsymbol{2},X) $ corresponds to the embedding  of the SM particles  and  $ (\boldsymbol{3},\boldsymbol{1},X) $, $ (\boldsymbol{1},\boldsymbol{1},X) $ is associated to particles beyond the SM.
Both possibilities $\boldsymbol{3}$ and $\boldsymbol{3^*}$ for $SU(3)_L$ multiplets are included in the flavor sector since the same number of 
fermion triplets and antitriplets must be present in order to cancel anomalies. One way to achieve this is choosing two quark families in one irreducible representation $\boldsymbol{3}$ ($\boldsymbol{3^*}$) and the other family of quarks and the three leptonic families in the representation $\boldsymbol{3^*}$ ($\boldsymbol{3}$). This way we guarantee a vector representation of fermions with  respect to the $SU(3)_L$ group, i.e. a model free of chiral anomalies.

Gauge bosons associated to the group $SU(3)_L$ transform according to the adjoint representation:
\begin{eqnarray}
\boldsymbol{W}_\mu&=&W_\mu^\alpha G_\alpha \nonumber\\
&=&\frac{1}{2}
\begin{bmatrix}
 W_\mu^3+\frac{1}{\sqrt{3}}W_\mu^8	& \sqrt{2}W_\mu^+ 			& \sqrt{2}K_\mu^{Q_1} \\
 \sqrt{2}W_\mu^- 			& -W_\mu^3+\frac{1}{\sqrt{3}}W_\mu^8 	& \sqrt{2}K_\mu^{Q_2}\\
 \sqrt{2}K_\mu^{-Q_1} 			& \sqrt{2}K_\mu^{-Q_2} 			& -\frac{2}{\sqrt{3}}W_\mu^8 &\\
\end{bmatrix}.
 \end{eqnarray}
The gauge field associated to $U(1)_X$ is represented as
\begin{equation}
 \boldsymbol{B}_\mu=\boldsymbol{I}_{3\times3}B_\mu.
\end{equation}
We have three gauge fields with $Q=0$ (that combine to form the photon and $Z$, $Z'$ bosons), two fields with $Q=\pm1$ ($W^\pm$)
and four fields with charges that depend on the choice of $\beta$ ($K_\mu^{Q_1}$, $K_\mu^{Q_2}$), with
\begin{eqnarray}
Q_1&=&\frac{1}{2}+\frac{\sqrt {3} \beta}{2},\nonumber \\
Q_2&=&-\frac{1}{2}+\frac{\sqrt {3} \beta}{2}.
\end{eqnarray}
Finally, the SSB follows the scheme
\begin{eqnarray*}
  {SU(3)_L\otimes U(1)_X}
 &&{\xrightarrow{ }}
 {SU(2)_L\otimes U(1)_Y} \\ 
&&{\xrightarrow{ }}
{U(1)_Q}.
\end{eqnarray*}
The transitions leave one massless gauge boson (photon), and eight massive (three weak and five exotic).
The generators that should be broken each time impose a restriction over the possible scalar fields.
Also, since we need Yukawa terms to give masses to the fermions, and such terms must be 331 invariant, 
we find that the scalar fields $\Phi$ can be in the representations $\boldsymbol{6}$, 
$\boldsymbol{3^*}$, or $\boldsymbol{3}$.

The choice of $\beta$, as well as the particle content (specially the scalar sector), and the possibility of imposing additional discrete symmetries, leads to a variety of 331 models. Since our object of study is the problem of neutrino masses, we will now briefly review some versions and how they explain them.


\subsection{Model $\boldsymbol {\beta=-\sqrt{3}}$, three Higgs triplets}
In \cite{Duong:1993zn, Montero:2001ji}, a model with three scalar triplets $\eta = \left(\eta^0, \eta^{-}_1 , \eta^+_2 \right)
 ^T \sim \left(1, 3, 0\right)$,
$\rho=\left(\rho^+, \rho^0, \rho^{++}\right)^T \sim \left(1, 3, 1\right)$ and
 $\chi = \left(\chi^-,\chi^{--},\chi^0\right)^T\sim \left(1, 3, -1\right)$, a leptonic triplet including  right-handed charged fields $\psi_{aL} =\left(\nu_a, l_a, l^C_a \right)_L \sim \left(1, 3, 0\right)$
and an additional heavy charged lepton singlet $E^{\prime}_L \sim \left(1, 1,−1\right)$ and $E ^\prime _R \sim \left(1, 1,−1\right)$ gives massive charged fields in the tree level, and neutrino mass matrix appears at the 1-loop level. With an extra neutral singlet $N_R$, there are new terms in the mass matrix from the tree-level.


\subsection{ Model $\boldsymbol {\beta=-\sqrt{3}}$, three Higgs triplets and a sextet}
 The model in \cite{Tully:2000kk} contains the same fields as the previous one (the three scalar triplets and the fermionic field $\psi_{aL}$), with the addition of a sextet 
\begin{equation} S=\left(
\begin{array}{ccc}
S^0_{11} & S^-_{12} & S^+_{13} \\
S^-_{12} & S^{--}_{22} & S^0_{23} \\
S^+_{13} & S^0_{23} & S^{++	}_{33} \\
\end{array}
\right),  \langle S \rangle = \left(
\begin{array}{ccc}
v ^\prime_S & 0 & 0 \\
0 & 0 &v_S \\
0 & v_S & 0 \\
\end{array}
\right).
 \end{equation}The Yukawa couplings with the fields $\eta$ and $S$ are the ones that give rise to Majorana mass terms; the ratio $v^\prime_S / v_S$ controls the relation between neutrino and charged lepton masses.


\subsection{Model $\boldsymbol {\beta=-1/\sqrt{3}}$ with two antisextets}
			      In the 331 model with $A_4$ flavour symmetry \cite{Dong:2010gk}, we have the leptons 
 $\psi_{aL}=(\nu_{aL},l_{aL},\nu^C_{aR})^T \sim(3,-1/3,\underline{3}),$
$e_{1R}\sim (1,-1,\underline{1})$, $ e_{2R}\sim (1,-1,\underline{1}')$, $
e_{3R}\sim (1,-1,\underline{1}''),$ and the scalar sector 
$\phi= \left( \phi^+_1, \phi^0_2, \phi^+_3 \right) ^T \sim (3,2/3,\underline{3}),$
$\eta= \left( \eta^0_1, \eta^-_2, \eta^0_3 \right) ^T \sim (3,-1/3,\underline{3}),$
$\chi= \left( \chi^0_1, \chi^-_2, \chi^0_3 \right) ^T \sim (3,-1/3,\underline{1}),$
$\rho= \left( \rho^+_1, \rho^0_2, \rho^+_3 \right) ^T \sim (3,2/3,\underline{1}),$
\begin{equation} S=\left(
\begin{array}{ccc}
S^0_{11} & S^-_{12} & S^0_{13} \\
S^-_{12} & S^{--}_{22} & S^-_{23} \\
S^0_{13} & S^-_{23} & S^0_{33} \\
\end{array}
\right), \qquad \langle S \rangle= \left(
\begin{array}{ccc}
\kappa_S & 0 & \vartheta_S \\
0 & 0 & 0 \\
\vartheta_S & 0 & \Lambda_S \\
\end{array}
\right),
 \end{equation}
where $S=s_1,\sigma$, and they transform as $s_1\sim (6^*,2/3,\underline{1})$,
$\sigma \sim (6^*,2/3,\underline{3})$,  under a
$(\mathrm{SU}(3)_L,\mathrm{U}(1)_X,A_4)$ symmetry. 
The charged leptons gain masses from the Yukawa interactions of the $SU(3)L$ triplet $\phi$, while 
quarks may gain masses either from $\phi$ (which leads to a CKM matrix equal to unity at first approximation), or
from $\eta$, $\chi$, $\rho$.
The Yukawa couplings with $s_1, \sigma$  lead to a neutrino mass matrix
\begin{equation} M_\nu\equiv\left(%
\begin{array}{cc}
  M_L & M^T_D \\
  M_D & M_R \\
\end{array}%
\right)\end{equation} where $M_{L,R,D}$ have the usual form, as does the
effective mass matrix for the active neutrinos
$M^{\mathrm{eff}}=M_L-M_D^TM_R^{-1}M_D$, which is a combination of type I
and type II seesaw mechanisms.


\subsection{Model $\boldsymbol {\beta=1/\sqrt{3}}$, four Higgs triplets and a singlet}
The model in \cite{Yin:2007rv} also considers the symmetry $A_4$, but with a different particle content. We have $\psi_{aL}=(\nu_{aL},l_{aL},E_{aL})^T \sim ( 3, -\frac{2}{3},\underline{3})
$, $N_{iR}\sim  (1,0,\underline{3}),$ $ e_{iR}\sim  (1,-1,\underline{1}^{(\ ,\prime,\prime\prime)}),$ $E_{iR}\sim  (1,-1,\underline{3})$, where $E_{iL,R}$ are negatively charged heavy leptons, and although the inclusion of right-handed neutral
Weyl states $N_{iR}$ is optional, they are used for the realization of tree-level canonical seesaw mechanism. In the scalar sector, we have 
 $\chi = (\chi^+, \chi^{\prime 0},\chi ^{0} )^T \sim (3, 1/3, \underline{1})$,
 $\eta=(\eta^0, \eta^{-}, \eta^{\prime -} )^T\sim ( 3, -2/3, \underline{1})$, 
$\eta_i=(\eta_i^0, \eta_i^{-}, \eta_i^{\prime -} )^T\sim (3, -2/3, \underline{3})$, 
$\rho =(\rho^+, \rho^0, \rho^{\prime 0} )^T \sim (3, 1/3, \underline{3})$,
$\xi \sim (1,0,\underline{3})$. The active neutrino mass matrix is $M^{\mathrm{eff}}=-M_D^TM_R^{-1}M_D$. With $M_N\sim TeV$, neutrinos will have mases $\sim 1eV$.


\subsection{Model $\boldsymbol {\beta=-1/\sqrt{3}}$, three Higgs triplets}
 In the model in \cite{ Dias:2005yh} we have the following relevant particle content:
 $\psi_{aL}=(\nu_{aL},l_{aL},\nu^c_{aR})^T \sim (3,
-1/3)$ $(a = 1, 2, 3)$, $l_{aR}\sim (1, -1)$. 
 $\chi = (\chi^0_1, \chi^-_2,
\chi^{0}_3 )^T \sim (3, -1/3)$, $\eta=(\eta^0_1, \eta^-_2,
\eta^{0}_3 )^T\sim (3, -1/3)$, $\rho =(\rho^+_1, \rho^0_2,
\rho^{+}_3 )^T \sim (3, 2/3)$, with the VEVs corresponding to
$\langle\chi \rangle = (0, 0,w/\sqrt{2})^T$, $\langle\eta
\rangle  = ( u/ \sqrt{2}, 0, 0)^T$, $\langle\rho \rangle =
(0, v/ \sqrt{2}, 0)^T$. 
The neutrinos do not achive mass in the tree level, but via radiative corrections.


\subsection{Model $\boldsymbol {\beta=-1/\sqrt{3}}$, three Higgs triplets and a sextet}
The models in \cite{Dong:2008sw, Cogollo:2008zc, Cogollo:2009yi} have the same content as above, with the addition of a scalar sextet
\begin{equation} S=\left(
\begin{array}{ccc}
S^0_{11} & S^-_{12} & S^0_{13} \\
S^-_{12} & S^{--}_{22} & S^-_{23} \\
S^0_{13} & S^-_{23} & S^0_{33} \\
\end{array}
\right), \qquad \langle S \rangle = \frac{1}{\sqrt{2}}\left(
\begin{array}{ccc}
\kappa & 0 & \vartheta \\
0 & 0 & 0 \\
\vartheta & 0 & \Lambda \\
\end{array}
\right).
 \end{equation}
This allows the obtention of tree-level mass terms, and leads to the effective mass for the active neutrinos 
 \begin{eqnarray}
M_1 &\simeq& -\sqrt{2}\left\{\left(\kappa
-\frac{\vartheta^2}{\Lambda}\right)
f^\nu-\frac{v^2}{\Lambda}h^\nu(f^\nu)^{-1}h^\nu\right\},\label{seesaw}\end{eqnarray}
where $f^\nu$ is the Yukawa coupling of the lepton triplets with the sextet, and $h^\nu$ the coupling with $\rho$.

\section{$\boldsymbol{\beta=-1/\sqrt{3}}$ model with RH singlet neutrinos \label{model}}
We consider a 331 model with $\beta=-\frac{1}{\sqrt{3}}$. The leptons are accommodated as follows: a triplet $l_{L}$, which includes the SM doublet in its first two entries and an exotic RH  neutrino $\nu _{R}^{C(i)}$ in the third; a RH charged lepton singlet $e_{R}$; and a RH neutral singlet $N_{R}$.
\begin{eqnarray}
l_{L}^{(i)}=%
\begin{pmatrix}
\nu _{L}^{(i)} \\ 
e_{L}^{(i)} \\ 
\nu _{R}^{C(i)}%
\end{pmatrix}%
& \sim \left( 1,3,-\frac{1}{3}\right) , \label{lepco1}\\
e_{R}^{(i)}& \sim \left( 1,1,-1\right) ,\label{lepco2} \\
N_{R}^{(i)}& \sim \left( 1,1,0\right) ,\label{lepco3}
\end{eqnarray}%
where the index $i=1,2,3$ represents the family (omitted from now on), and the symbol $\sim$ refers to the representation under the $SU(3)_C \otimes SU (3)_L \otimes U(1)_X$ gauge group.
For the quarks, we have 
\begin{eqnarray}
q_{L}^{(m)}=%
\begin{pmatrix}
d _{L}^{(m)} \\ 
-u_{L}^{(m)} \\ 
D_{L}^{(m)}%
\end{pmatrix}%
& \sim \left( 3, 3^*,0\right) , \label{quarco1} \\
q_{L}^{3}=%
\begin{pmatrix}
u_{L}^{(3)} \\ 
d_{L}^{(3)} \\ 
U_{L}%
\end{pmatrix}%
& \sim \left( 3, 3,\frac{1}{3}\right) , \label{quarco2}\\
u_{R}^{(i)}& \sim \left(3,1,\frac{2}{3}\right) ,\label{quarco3} \\
d_{R}^{(i)}& \sim \left(3,1,-\frac{1}{3}\right) ,\label{quarco4} \\
U_{R}& \sim \left(3,1,\frac{2}{3}\right) ,\label{quarco5} \\
D_{R}^{(m)}& \sim \left(3,1,-\frac{1}{3}\right), \label{quarco6}
\end{eqnarray}
with $m=1,2$.
The fermionic content (\ref{lepco1}-\ref{quarco6}) is anomaly free \cite{Martinez:2006gb, Diaz:2004fs}. From here on, we will focus exclusively on the leptonic sector.
For the scalar sector, we have three triplets
\begin{eqnarray}
\chi &=&%
\begin{pmatrix}
\chi _{1}^{0} \\ 
\chi _{2}^{-} \\ 
\frac{v_{\chi }+\xi \chi +i\ \zeta _{\chi }}{\sqrt{2}}%
\end{pmatrix}%
\sim \left( 1,3,-\frac{1}{3}\right) ,  \nonumber \\
\eta &=&\left( 
\begin{array}{c}
\frac{v_{\eta }+\xi \eta +i\ \zeta _{\eta }}{\sqrt{2}} \\ 
\eta _{2}^{-} \\ 
\eta _{3}^{0}%
\end{array}%
\right) \sim \left( 1,3,-\frac{1}{3}\right) ,  \nonumber \\
\rho &=&\left( 
\begin{array}{c}
\rho _{1}^{+} \\ 
\frac{v_{\rho }+\xi \rho +i\ \zeta _{\rho }}{\sqrt{2}} \\ 
\rho _{3}^{+}%
\end{array}%
\right) \sim \left( 1,3,\frac{2}{3}\right) .
\end{eqnarray}%
The SSB follows the scheme
$
  {SU(3)_L\otimes U(1)_X}
 {\xrightarrow{\langle \chi \rangle}}
 {SU(2)_L\otimes U(1)_Y} 
{\xrightarrow{\langle \eta \rangle,\langle \rho \rangle}}
{U(1)_Q},
$
where the vacuum expectation values satisfy $v_{\chi }\gg
v_{\eta },v_{\rho }.$ 

The most general potential that we can construct with three scalar triplets is:
\begin{eqnarray}
&V_{H} &=\mu _{\chi }^{2}(\chi ^{\dagger }\chi )+\mu _{\eta }^{2}(\eta
^{\dagger }\eta )+\mu _{\rho }^{2}(\rho ^{\dagger }\rho )
+f  \left(
\chi _{i}\eta _{j}\rho _{k}\varepsilon ^{ijk}+H.c.\right)   
+\lambda _{1}(\chi ^{\dagger }\chi )(\chi ^{\dagger }\chi )\nonumber \\ &&
+\lambda _{2}(\rho ^{\dagger }\rho )(\rho ^{\dagger }\rho )
+\lambda _{3}(\eta ^{\dagger }\eta )(\eta ^{\dagger }\eta ) 
+\lambda _{4}(\chi ^{\dagger }\chi )(\rho ^{\dagger }\rho )  
+\lambda _{5}(\chi ^{\dagger }\chi )(\eta ^{\dagger }\eta )\nonumber \\ &&
+\lambda _{6}(\rho^{\dagger }\rho )(\eta ^{\dagger }\eta )  
+\lambda _{7}(\chi ^{\dagger }\eta )(\eta ^{\dagger }\chi )
+\lambda_{8}(\chi ^{\dagger }\rho )(\rho ^{\dagger }\chi )
+\lambda _{9}(\rho^{\dagger }\eta )(\eta ^{\dagger }\rho ). \label{v00}
\end{eqnarray}%
For $\beta=-\frac{1}{\sqrt{3}}$ we could write the additional terms
\begin{eqnarray}
V_{-1/{\sqrt{3}}}&=&\mu_4^2 (\chi^{\dagger}\eta + H.c.) 
+\lambda _{10}(\chi ^{\dagger }\chi )(\chi^{\dagger}\eta + H.c.)
+\lambda _{11}(\eta ^{\dagger }\eta)(\eta^{\dagger}\chi + H.c.)  \nonumber \\ &&
+\lambda _{12}(\rho^{\dagger }\rho )(\chi^{\dagger}\eta + H.c.)
+\lambda _{13}(\chi^{\dagger }\eta \chi^{\dagger}\eta + H.c.)
 +\lambda _{14}(\rho^{\dagger }\chi \eta^{\dagger}\rho + H.c.) . \label{v13}
\end{eqnarray}%
Although those additional terms would modify the mass matrices in the scalar spectrum introducing the
new parameters $\mu_4,\lambda_{10-14},$ they do not add to the predictability of the model, in the sense that we would still have the same number of Higgs and Goldstone bosons with similar mass and mixing structures, but with an enlarged number of variables to fit.  An elegant mechanism to get rid of terms in  (\ref{v13}) would be to introduce a discrete symmetry $\xi \rightarrow \xi,$ $\eta \rightarrow -\eta,$ $\rho \rightarrow \rho,$ plus the condition $\lambda_{13}=0$; however, it leads to a poorer Yukawa sector which is defeating to the goal of this work and thus will not be implemented. For the sake of simplicity, we only present the mass eigenstates for the potential  (\ref{v00})  in table \ref{scalart} \cite{Diaz:2003dk}, keeping in mind that adding (\ref{v13}) does not have significant implications for the main results of following sections.


 \begin{table*}
 \caption{\label{scalart} Mass eigenstates of the scalar sector in the 331 model described. $\tan \beta =\frac{v_{\rho }}{v_{\eta }}$}
 \begin{ruledtabular}
 \begin{tabular}{llll} 
$H_3^0=\xi_\chi$ &$ M^2_{H^3_0}=4 \lambda_1 v_\chi^2 $ \\
$h^{0}=\cos  \alpha  \xi _{\eta }+\sin  \alpha 
\xi _{\rho }$ & $M_{h^{0}}^{2}=\frac{4\left( \lambda _{3}v_{\eta
}^{4}+v_{\rho }^{2}\lambda _{6}v_{\eta }^{2}+v_{\rho }^{4}\lambda
_{2}\right) }{v_{\eta }^{2}+v_{\rho }^{2}}$ & $G_{3}^{0}\approx -\zeta
_{\chi }$ & $M_{G_{3}^{0}}^{2}=0$ \\ 
$H^{0}=-\sin  \alpha  \xi _{\eta }+\cos  \alpha 
\xi _{\rho }$ & $M_{H^{0}}^{2}=-\frac{\sqrt{2}f\left( v_{\eta }^{2}+v_{\rho
}^{2}\right) v_{\chi }}{v_{\eta }v_{\rho }}$ & $G^{0}\approx \cos \beta
\zeta _{\eta }-\sin \beta \zeta _{\rho }$ & $M_{G^{0}}^{2}=0$ \\ 
$A^{0}\approx \sin \beta \zeta _{\eta }-\cos \beta \zeta _{\rho }$ & $%
M_{A^{0}}^{2}\approx -\sqrt{2}fv_{\chi }\left( \frac{v_{\rho }}{v_{\eta }}+%
\frac{v_{\eta }}{v_{\rho }}\right) $ & $G_{1}^{0}\approx -\chi _{1}^{0}$ & $%
M_{G_{1}^{0}}^{2}=0$ \\ 
$H_{1}^{0}\approx \eta _{3}^{0}$ & $M_{H_{1}^{0}}^{2}=\frac{-1}{\sqrt{2}}%
fv_{\rho }\left( \frac{v_{\eta }}{v_{\chi }}+\frac{v_{\chi }}{v_{\eta }}%
\right) +\lambda _{7}\frac{\left( v_{\eta }^{2}+v_{\chi }^{2}\right) }{2}$ & 
$G_{2}^{\pm }\approx -\chi _{2}^{\pm },$ & $M_{G_{2}^{\pm }}^{2}=0$ \\ 
$H_{2}^{\pm }\approx \rho _{3}^{\pm }$ & $M_{H_{2}^{\pm }}^{2}=\frac{-1}{%
\sqrt{2}}fv_{\eta }\left( \frac{v_{\rho }}{v_{\chi }}+\frac{v_{\chi }}{%
v_{\rho }}\right) +\lambda _{8}\frac{\left( v_{\rho }^{2}+v_{\chi
}^{2}\right) }{2}$ & $G^{\pm }=-\cos \beta \eta _{2}^{\pm }+\sin \beta \rho
_{1}^{\pm }$ & $M_{G^{\pm }}^{2}=0$ \\ 
$H^{\pm }=\sin \beta \eta _{2}^{\pm }+\cos \beta \rho _{1}^{\pm }.$ & $%
M_{H^{\pm }}^{2}=\frac{-1}{\sqrt{2}}fv_{\chi }\left( \frac{v_{\rho }}{%
v_{\eta }}+\frac{v_{\eta }}{v_{\rho }}\right) +\lambda _{9}\frac{\left(
v_{\rho }^{2}+v_{\eta }^{2}\right) }{2}$ &  & 
\end{tabular}
 \end{ruledtabular}
 \end{table*}

The relevant part of the Lagrangian constructed with these fields (Yukawa Lagrangian plus a Majorana mass term) is
\begin{eqnarray}
-\mathcal{L} & \supset
& h_{\rho e}\bar{l}\rho e_{R}+h_{\chi }\bar{l}.\chi
N_{R}+h_{\eta }\bar{l}\eta N_{R}+ \nonumber \\
&& \frac{1}{2}h_{\rho }\left( \overline{l_{L}}%
\right) ^{a}\left( l_{L}^{C}\right) ^{b}\rho ^{c}\varepsilon _{abc}+\frac{1}{%
2}M_{R}\overline{N_{R}}N_{R}^{C}+H.c.,
\end{eqnarray}%

Using the VEV of the fields, we find the mass Lagrangian for the leptonic fields
\begin{eqnarray}
-\mathcal{L}_{\text{mass}}^{l} &=&\frac{v_{\rho }}{\sqrt{2}}\overline{e_{L}}h_{\rho
e}e_{R}-\frac{1}{2\sqrt{2}}v_{\rho }\overline{\nu _{L}}h_{\rho }\nu _{R}+%
\frac{1}{2\sqrt{2}}v_{\rho }\overline{\nu _{R}^{C}}h_{\rho }\nu _{L}^{C} \nonumber \\
&&+\frac{1}{\sqrt{2}}v_{\chi }\overline{\nu _{R}^{C}}h_{\chi }N_{R}+\frac{1}{%
\sqrt{2}}v_{\eta }\overline{\nu _{L}}h_{\eta }N_{R}+
\frac{1}{2}M_{R}%
\overline{N_{R}}N_{R}^{C}+H.c. \label{lagmass}
\end{eqnarray}


\subsection{Neutrino mass matrices}
Rearranging the terms involving neutrinos in (\ref{lagmass}), we can write for the neutrinos
\begin{equation}
-\mathcal{L}_{\text{mass}}^{\nu }=\frac{1}{2}\left( 
\begin{array}{ccc}
\overline{\nu _{L}^{C}} & \overline{\nu _{R}} & \overline{N_{R}}%
\end{array}%
\right) M
\begin{pmatrix}
\nu _{L} \\ 
\nu _{R}^{C} \\ 
N_{R}^{C}%
\end{pmatrix}%
+H.c.,
\end{equation}%
with
\begin{equation}
M=\left( 
\begin{array}{ccc}
0 & \frac{v_{\rho }h_{\rho }^{\prime \ast }}{\sqrt{2}} & \frac{v_{\eta
}h_{\eta }^{\ast }}{\sqrt{2}} \\ 
\frac{v_{\rho }h_{\rho }^{\prime ^{\dag }}}{\sqrt{2}} & 0 & \frac{v_{\chi
}h_{\chi }^{\ast }}{\sqrt{2}} \\ 
\frac{v_{\eta }h_{\eta }^{\dagger }}{\sqrt{2}} & \frac{v_{\chi }h_{\chi
}^{\dagger }}{\sqrt{2}} & M_{R}%
\end{array}%
\right)   \label{neutrinomassmatrix},
\end{equation}
where we have defined 
$h_{\rho }^{\prime }=\frac{h_{\rho }^{T}-h_{\rho }}{2}
$.
This mass matrix has a similar structure to some worked previously in the literature \cite{Barr:2003nn,Ibarra:2010xw}, considering the restriction $v_\chi \gg v_\rho, v_\eta$. Notice that the Majorana mass, $M_R$, is non-restricted so far; it could be in a smaller or larger scale than the VEV's. Nevertheless, 
the diagonalization of the matrix $M$ as presented leads in first approximation to a light neutrino mass
\begin{eqnarray}
M_{\xi1} &=&-\frac{v_{\eta }v_{\rho }}{\sqrt{2}v_{\chi }}\left(
h_{\rho }^{\prime \ast }\left( h_{\chi }^{\dagger }\right) ^{-1}h_{\eta }^{\dag
}+h_{\eta }^{\ast }\left( h_{\chi }^{\dagger }\right) ^{-1}h_{\rho }^{\prime\dag
}\right) \nonumber\\ && +\frac{v_{\rho }^{2}}{v_{\chi} ^{2}}h_{\rho }^{\prime\ast }\left( h_{\chi
}^{\dagger }\right) ^{-1}M_R \left( h_{\chi }^{\dagger }\right) ^{-1}h_{\rho
}^{\prime\dag }. \label{general}
\end{eqnarray}

Notice that both terms depend on the antisymmetric matrix $h_\rho^\prime$, therefore, the symmetries in the Lagrangian must be such that the Yukawa coupling term between 
 $\nu_L$ and $N_R$ does not cancel. Furthermore, it should not be symmetric. On the other hand, we can set to zero the 13 entry of the mass matrix by using a discrete symmetry and we would still have tree-level masses for the lightest neutrinos.

The matrix $M_R$ sets a scale for the breaking of  lepton number; if $M_R \ll v_\rho, v_\chi, v_\eta,$ then the fields $\eta_3^0$ and 
$\chi_1^0$ could acquire VEVs \cite{Dong:2010gk}, say, $v_\eta^\prime,$ $v_\chi^\prime$. Their values are restricted since they are also involved in the  mixing among the exotic quarks and
ordinary quarks of the same charge, thus in flavor-changing neutral-current processes; to keep a consistency with the effective theory, it is safe to impose the constraints  $v_\eta^\prime \ll  v_\chi,$ $v_\chi^\prime \ll  v_\eta$ \cite{Dong:2008sw}. Now, for neutrino masses, to include these terms
we would have to do the substitutions %
$
v_{\chi }h_{\chi }^{\dagger } \rightarrow v_{\chi }h_{\chi }^{\dagger
}+v_{\eta }^{\prime }h_{\eta }^{\dagger }, $ $
v_{\eta }h_{\eta }^{\dagger } \rightarrow v_{\eta }h_{\eta }^{\dagger
}+v_{\chi }^{\prime }h_{\chi }^{\dagger }$ in (\ref{general}).
Since we expect Yukawa matrices to be roughly of the same order, in cases
where we keep the terms with both $h_{\chi}$ and $h_{\eta }$ (such as cases 1 and 3 below), 
we can safeley neglect the extra terms. Now, if some extra symmetry cancels  the terms with 
$h_{\eta }$ (e.g. case 2 below), we would only have the substitution $%
v_{\eta }h_{\eta }^{\dagger }\rightarrow v_{\chi }^{\prime }h_{\chi
}^{\dagger };$ then, instead of the first term in (\ref{general}), we would have $-\frac{v_{\chi }^{\prime }v_{\rho }}{%
\sqrt{2}v_{\chi }}\left( h_{\rho }^{\prime \ast }\left( h_{\chi }^{\dagger
}\right) ^{-1}h_{\chi }^{\dagger }+h_{\chi }^{\ast }\left( h_{\chi
}^{\dagger }\right) ^{-1}h_{\rho }^{\prime \dagger }\right) 
= -\frac{v_{\chi }^{\prime }v_{\rho }}{\sqrt{2}%
v_{\chi }}\left( h_{\rho }^{\prime \ast }+h_{\rho }^{\prime \dagger }\right), $
which exactly cancels since $h_{\rho }^{\prime }$ is antisymmetric. Therefore, we can safely
assume that the presence of the small expectation
values does not affect the results for neutrino masses in any of the cases
considered.

The complete diagonalization procedure, with the rotation matrices and mass eigensystems, is presented in \ref{dianeu} using the schemes of appendix \ref{diag}. 
For simplicity, we will consider the following scenarios:
\begin{enumerate}
	\item   $M_R \gg  v_\chi \gg v_\eta,v_\rho $. This is the \textit{double seesaw} mechanism \cite{Barr:2003nn, Gu:2010xc}. We have three different mass scales for neutrinos: very light active neutrinos ($\xi_1$), and heavy ($\xi_2$) and \textit{very} heavy ($\xi_3$) sterile neutrinos:
\begin{subequations}
\begin{eqnarray}
M_{\xi_1} &=&\frac{v_{\rho }^{2}}{v_{\chi} ^{2}}h_{\rho }^{\prime\ast }\left( h_{\chi
}^{\dagger }\right) ^{-1} M_R \left( h_{\chi }^{\dagger }\right) ^{-1}h_{\rho
}^{\prime\dag } ,\\
M_{\xi_2} &=& -\frac{v_\chi ^2}{2} h_{\chi}^{\ast} M_{R}^{-1} h_{\chi }^{\dagger }, \\
M_{\xi_3} &=&M_R.
\end{eqnarray}
\end{subequations}
Since the only constraints that we have are $v_\chi  \gtrsim 10^3$ GeV, $v_\eta^2 + v_\rho^2 = v^2 \sim 10^2$ GeV,
we may set $v_\chi  \sim 10^4$, $h_\chi \sim 1$, $M_R\sim 10^6$,  $v_\rho h_\rho^\prime \sim 10^{-4}$ and find $M_{\xi_1}\sim 0.1$ eV,$M_{\xi_2}\sim 10^2$ GeV,$M_{\xi_3}\sim 10 ^6$ GeV. The $\xi_2$ neutrinos are candidates for detection in LHC, while $\xi_3$  create a scenario for the study leptogenesis.
\item $M_R=\mu \ll v_\eta,v_\rho \ll v_\chi$ and an additional discrete symmetry which cancels the 13 and 31 entries. 
\begin{subequations}
\begin{eqnarray}
M_{\xi_1} &=&\frac{v_{\rho }^{2}}{v_{\chi} ^{2}}h_{\rho }^{\prime\ast }\left( h_{\chi
}^{\dagger }\right) ^{-1}\mu \left( h_{\chi }^{\dagger }\right) ^{-1}h_{\rho
}^{\prime\dag } ,\\
M_{\xi_2} &=& -\frac{v_\chi }{\sqrt{2}} h_{\chi}^{\dagger } + \frac{\mu}{2}, \\
M_{\xi_3} &=& \frac{v_\chi }{\sqrt{2}} h_{\chi}^{\dagger } + \frac{\mu}{2}.
\end{eqnarray}
\end{subequations}
This is the \textit{inverse seesaw} mechanism \cite{Mohapatra:1986aw, Mohapatra:1986bd, Gu:2010xc}. The exotic neutrinos are pseudo-Dirac, with masses $\sim \pm \frac{v_\chi h_\chi}{\sqrt{2}} $ and a small splitting $\sim \mu$.
There are several ways to obtain sub-eV masses for the active neutrinos. For example, a set of parameters that works and is consistent is   $v_\chi  \sim 10^4$ , $ h_\chi \sim 1$, $\mu \sim 10^{-6}$,   $v_\rho h_\rho^\prime \sim 1$ and leads to $M_{\xi_1}\sim 0.1$ eV,$M_{\xi_2}\sim 10^4$ GeV,$M_{\xi_3}\sim 10 ^4$ GeV. All the exotic neutrinos are in the TeV scale.
\item $M_R =0$. Even without the Majorana mass term, we still have mass matrices for the neutrinos, that depend on the known scales $v_\chi \sim TeV$, $v_\eta^2+v_\rho^2 = v^2$, in the \textit{linear seesaw} mechanism  \cite{Gu:2010xc} . 
\begin{subequations}
\begin{eqnarray}
M_{\xi_1} &=&-\frac{v_{\eta }v_{\rho }}{\sqrt{2}v_{\chi }}\left(
h_{\rho }^{\prime \ast }\left( h_{\chi }^{\dagger }\right) ^{-1}h_{\eta }^{\dag
}+h_{\eta }^{\ast }\left( h_{\chi }^{\dagger }\right) ^{-1}h_{\rho }^{\prime\dag
}\right), \\
M_{\xi_2} &=& -\frac{v_\chi }{\sqrt{2}} h_{\chi}^{\dagger } , \\
M_{\xi_3} &=& \frac{v_\chi }{\sqrt{2}} h_{\chi}^{\dagger } .
\end{eqnarray}
\end{subequations}
A possible set of parameters is  $v_\chi  \sim 10^4$ GeV, $ h_\chi \sim 1$,  $v_\eta  \sim 1$GeV,  $v_\rho  \sim 10^2$GeV, $h_\rho \sim h_\eta \sim 10^{-4}$.
\end{enumerate}

From the discussion above we see that it is indeed possible to get the right orders of magnitude for light neutrino masses from (\ref{general}). However,  $M_{\xi 1}$  is presented as the product of $3\times 3$ complex matrices $h_\rho^\prime, $ $h_\chi$, $h_\eta$, $M_R$, and the only restriction we have on their structure so far is that $h_\rho^\prime$ must be antisymmetric. Furthermore, they are independent of the charged lepton mass matrix ($\propto h_{\rho e}$ as seen in \ref{lagmass}).  The problem of choosing ansatz that lead to neutrino masses and mixing compatible with experimental data has been treated extensively \cite{King:2002nf, * Strumia:2006db,* Grimus:2004hf, * Altarelli:2007gb}, and the specific application to our model with a proper scan of the parameter space will be treated elsewhere.

\subsection{Diagonalization of the neutrino mass matrix \label{dianeu}}
To put the neutrino mass matrix (\ref{neutrinomassmatrix}) in diagonal form, we block-diagonalize using multiple times  the procedures in \ref{dia1}, \ref{dia2}, and then use the PMNS matrix for active neutrinos, and define PMNS-like matrices for the exotic neutrinos. 
\subsubsection{Large $M_R$}
For large $M_R$, i.e. the double seesaw mechanism, we use   \ref{dia1} three times.
First, we combine the 2,3 entries with the rotation matrix $W_1$. Then, the 1,3 entries with $W_2$, and finally the 1,2 entries with $W_3$. The rotation matrices are 
\begin{eqnarray}
W_1 &=& \left( 
\begin{array}{ccc}
1 & 0 & 0 \\ 
0 & 1 & B_{1} \\ 
0 & -B_{1}^{\dagger } & 1
\end{array}%
\right) ,\ \ 
W_2=\left( 
\begin{array}{ccc}
1 & 0 & B_{2} \\ 
0 & 1 & 0 \\ 
-B_{2}^{\dagger } & 0 & 1%
\end{array}%
\right) , \ \ 
W_3=\left( 
\begin{array}{ccc}
1 & B_{3} & 0 \\ 
-B_{3}^{\dagger } & 1 & 0 \\ 
0 & 0 & 1%
\end{array}%
\right) ,
\end{eqnarray}
with
\begin{subequations}
\begin{eqnarray}
B_{1}^{\dag }& =&M_{R}^{-1}\frac{v_{\chi }h_{\chi }^{\dagger}}{\sqrt{2}}, \qquad \qquad
B_{2}^{\dag } = M_{R}^{-1} \frac{v_{\eta }h_{\eta }^{\dagger }}{\sqrt{2}},\\
B_{3}^{\dag }&=&-\frac{\sqrt{2}v_\rho}{v_\chi^2} h_{\rho} ^{\prime \ast} \left(  h_{\chi}^{\dagger} \right) ^{-1}
M_R  \left(  h_{\chi}^{\ast} \right) ^{-1} + \frac{v_\eta}{v_\chi}  \left(  h_{\chi}^{\dagger} \right) ^{-1}   h_{\eta} ^{ \dagger}.
\end{eqnarray}\end{subequations}
The product of these rotations with the PMNS-like matrices, up to first order in the $B_i$'s, is
\begin{eqnarray}
\mathbb{U}=
W_1 \cdot W_2  \cdot W_3 \cdot \left( 
\begin{array}{ccc}
V & 0 & 0 \\ 
0 & U_{\chi } & 0 \\ 
0 & 0 & U_{R}%
\end{array}%
\right) =
\begin{pmatrix}
V & B_{3}U_{\chi } & B_{2}U_{R} \\ 
-B_{3}^{\dagger }V & U_{\chi } & B_{1}U_{R} \\ 
-B_{2}^{\dagger }V & -B_{1}^{\dagger }U_{\chi } & U_{R}%
\end{pmatrix}.%
\end{eqnarray}
The mass eigenstates $\xi$ are constructed as 
\begin{equation}
n_L = \mathbb{U} \xi_L, \qquad n_L^C = \mathbb{U}^{\ast} \xi_R,
\end{equation}
with $n_L=\left(   \nu_L, \nu_R ^C, N_R^C \right)^T$, $\xi=\left( \xi_1,\xi_2, \xi_3 \right) ^T$, and the mass matrix
\begin{equation} \label{blockmass}
M_{\xi}^{\text{diag}}=\mathbb{U}^T M \mathbb{U}
\begin{pmatrix} V^{T}  M_{\xi_1} V &0 & 0\\ 0 &  U_R^{T}  M_{\xi_2} U_R & 0\\ 0&0  & U_\chi^{T}  M_{\xi_3} U_\chi  \end{pmatrix},
\end{equation} where
\begin{subequations}
\begin{eqnarray}
M_{\xi_1} &=&\frac{v_{\rho }^{2}}{v_{\chi} ^{2}}h_{\rho }^{\prime\ast }\left( h_{\chi
}^{\dagger }\right) ^{-1} M_R \left( h_{\chi }^{\dagger }\right) ^{-1}h_{\rho
}^{\prime\dag }
-\frac{v_\eta v_\rho}{\sqrt{2} v_\chi} \left(
h_{\rho }^{\prime\ast } \left( h_{\chi}^{\dagger }\right)  ^{-1}   h_{\eta}^{\dagger}+
h_{\eta }^{\ast } \left( h_{\chi}^{\dagger }\right)  ^{-1}   h_{\rho}^{\prime \dagger}
              \right),
 \\
M_{\xi_2} &=& -\frac{v_\chi ^2}{2} h_{\chi}^{\ast} M_{R}^{-1} h_{\chi }^{\dagger } , \qquad \qquad \qquad
M_{\xi_3} =M_R.
\end{eqnarray}
\end{subequations}

Notice that the neutrino fields thus constructed are Majorana: $\xi= \xi_L + \xi_R = \mathbb{U}^\dagger n_L +  \left( \mathbb{U}^\dagger n_L\right)^C = \xi ^C.$


\subsubsection{Small $M_R$}
For small $M_R$ ($M_R =\mu$), we have to use the diagonalization \ref{dia2} for the 2,3 entries, and twice \ref{dia1} for the 1,3 and 1,2 entries. The rotation matrices are 
\begin{eqnarray}
W_1 &=& \left( \begin{array}{ccc}1 & 0 & 0 \\ 0 & \frac{ (1-S)}{\sqrt{2} }& \frac{ (1+S)}{ \sqrt{2}} \\ 0 & \frac{ (-1-S)}{ \sqrt{2}} & \frac{ (1-S)}{ \sqrt{2}}
\end{array}%
\right) ,\ \ 
W_2=\left( \begin{array}{ccc}1 & 0 & B_{2} \\ 0 & 1 & 0 \\ -B_{2}^{\dagger } & 0 & 1%
\end{array}%
\right) , \ \ 
W_3=\left( \begin{array}{ccc}1 & B_{3} & 0 \\ -B_{3}^{\dagger } & 1 & 0 \\ 0 & 0 & 1%
\end{array}%
\right) ,
\end{eqnarray}
with
\begin{subequations}
\begin{eqnarray}
S &=&-\frac{1}{2\sqrt{2}v_{\chi }} \left( h_{\chi }^{\dag }\right) ^{-1}\mu, \\
B_2 ^\dagger &=& \frac{v_\eta}{\sqrt{2} v_\chi} \left( h_\chi ^ {\dagger}\right) ^{-1} h_\eta ^{\dagger}
+ \frac{v_\rho}{\sqrt{2} v_\chi} \left( h_\chi ^ {\dagger}\right) ^{-1} h_\rho ^{\prime \dagger}
+  \left( h_\chi ^ {\dagger}\right) ^{-1} \mu \left( h_\chi ^ {\dagger}\right) ^{-1}
 \left( \frac{-v_\eta}{4 v_\chi^2} h_\eta^{\dagger}   + \frac{v_\rho}{4 v_\chi^2} h_\rho^{\prime\dagger}             \right), \\
B_3 ^\dagger &=& \frac{v_\eta}{\sqrt{2} v_\chi} \left( h_\chi ^ {\dagger}\right) ^{-1} h_\eta ^{\dagger}
+ \frac{v_\rho}{\sqrt{2} v_\chi} \left( h_\chi ^ {\dagger}\right) ^{-1} h_\rho ^{\prime \dagger}
+  \left( h_\chi ^ {\dagger}\right) ^{-1} \mu \left( h_\chi ^ {\dagger}\right) ^{-1}
 \left( \frac{v_\eta}{4 v_\chi^2} h_\eta^{\dagger}   - \frac{3v_\rho}{4 v_\chi^2} h_\rho^{\prime\dagger}             \right).
\end{eqnarray}
\end{subequations}
The full rotation matrix, up to first order in the $S$,$B_i$'s, is
\begin{eqnarray}
\mathbb{U}=
W_1 \cdot W_2  \cdot W_3 \cdot \left( 
\begin{array}{ccc}
V & 0 & 0 \\ 
0 & U_{\chi } & 0 \\ 
0 & 0 & U_{R}%
\end{array}%
\right) =
\begin{pmatrix}
V & B_3 U_\chi & B_2 U_R \\
-\frac{(B_2^\dagger+B_3^\dagger)}{\sqrt{2}} V&   \frac{(1-S)}{\sqrt{2}} U_\chi  &   \frac{(1+S)}{\sqrt{2}} U_R \\
-\frac{(B_2^\dagger-B_3^\dagger)}{\sqrt{2}} V&   \frac{(-1-S)}{\sqrt{2}} U_\chi  &   \frac{(1-S)}{\sqrt{2}} U_R
\end{pmatrix},%
\end{eqnarray}
and the blocks in the mass matrix (\ref{blockmass})
\begin{subequations}
\begin{eqnarray}
M_{\xi_1} &=&\frac{v_{\rho }^{2}}{v_{\chi} ^{2}}h_{\rho }^{\prime\ast }\left( h_{\chi
}^{\dagger }\right) ^{-1}\mu \left( h_{\chi }^{\dagger }\right) ^{-1}h_{\rho
}^{\prime\dag }-\frac{v_{\eta }v_{\rho }}{\sqrt{2}v_{\chi }}\left(
h_{\rho }^{\prime \ast }\left( h_{\chi }^{\dagger }\right) ^{-1}h_{\eta }^{\dag
}+h_{\eta }^{\ast }\left( h_{\chi }^{\dagger }\right) ^{-1}h_{\rho }^{\prime\dag
}\right), \\
M_{\xi_2} &=& -\frac{v_\chi }{\sqrt{2}} h_{\chi}^{\dagger } + \frac{\mu}{2} ,\\
M_{\xi_3} &=& \frac{v_\chi }{\sqrt{2}} h_{\chi}^{\dagger } + \frac{\mu}{2}.
\end{eqnarray}
\end{subequations}


\section {Conclusion}
In this paper we have discussed an extension to the Standard Model using the $SU(3)_C \otimes SU(3)_L \otimes U(1)_X$ ($331$) symmetry group. This symmetry is spontaneously broken to the SM group at the TeV scale. In a model with $\beta = -1/\sqrt{3}$, with three scalar triplets, and one lepton triplet (one charged, two neutral fields) and two singlets (one charged, one neutral) for each family, we find that neutrinos may acquire tiny masses via the double, inverse or linear seesaw mechanisms. According to the selection of the Majorana mass term $M_R$ for the singlet neutrino and additional discrete symmetries, we have three different scenarios for the exotic neutrinos: for large $M_R$, three sterile neutrinos ($\xi_2$) are in an intermediate scale (e.g. $\sim 10^{2,3}$ GeV), and other three ($\xi_3$) in a heavier scale ($\sim 10^{6,7}$ GeV); for small $M_R$ we have three pairs of pseudo-Dirac neutrinos in the TeV scale; and for $M_R=0$ they combine to form three Dirac neutrinos with masses in the TeV scale as well.
The model presented gives a mechanism for tree-level neutrino masses in a rather simple extension of the SM, without the introduction of doubly-charged Higgs particles or any other kind of exotic charges.


\begin{acknowledgments}
The authors would like to thank Colciencias and the Mazda Foundation for the Arts and Sciences for financial support.
\end{acknowledgments}


\appendix
\section{\label{diag}Diagonalization of mass matrices - General case}

We consider a unitary matrix $\Omega $ which block-diagonalizes the mass matrix $M$
\begin{equation}
\Omega ^{T} M \Omega =
\Omega ^{T}\left( 
\begin{array}{cc}
0 & M_{D} \\ 
M_{D}^{T} & M_{N}%
\end{array}%
\right) \Omega =\left( 
\begin{array}{cc}
U^{\ast }m^{\text{diag}}U^{\dagger } & \mathbb{O} \\ 
\mathbb{O}^{T} & V^{\ast }M^{\text{diag}}V^{\dagger }%
\end{array}%
\right) \label{diag00}
\end{equation}%
with $\mathbb{O}$ a matrix with all null entries.
 $\Omega $ may be formally expressed as the exponential of a anti-hermitian matrix \cite{Ibarra:2010xw}
\begin{equation}
\Omega =\exp  \left( 
\begin{array}{cc}
\mathbb{O} & R \\ 
-R^{\dagger } & \mathbb{O}%
\end{array}%
\right) .
\end{equation}


\subsection{ For $M_{N}$ $\gg $ $M_{D}$ \label{dia1}}
If $M_{N}$ $\gg $ $M_{D}$ (e.g. the Majorana mass of a RH neutrino belongs to a heavy scale)
\begin{equation}
\Omega  =\left( 
\begin{array}{cc}
1-\frac{RR^{\dagger }}{2} & R \\ 
-R^{\dagger } & 1-\frac{R^{\dagger }R}{2}%
\end{array}%
\right) +O\left( R^{3}\right).
\end{equation}
Using (\ref{diag00}), from the off-diagonal elements we find
\begin{eqnarray}
R^{\ast } =M_{D}M_{N}^{-1}, \qquad
R^{\dagger } =M_{N}^{-1}M_{D}^{T},
\end{eqnarray}%
 And substituting for the diagonal elements, we get the mass matrices%
\begin{eqnarray}
U^{\ast }m^{\text{diag}}U^{\dagger } &=&-R^{\ast }M_{D}^{T} 
=-M_{D}M_{N}^{-1}M_{D}^{T}  \nonumber \\
&=&-R^{\ast }M_{N}R^{\dagger },\\
V^{\ast }M^{\text{diag}}V^{\dagger }
&=&M_{N}+\frac{1}{2}(R^{T}M_{D}+M_{D}^{T}R).
\end{eqnarray}


\subsection{For $M_{N}\ll M_{D}$  \label{dia2}}
When $M_{N}\ll M_{D}$ (for example, if neutrinos are pseudo-Dirac)
\begin{eqnarray}
\Omega =\frac{1}{\sqrt{2}}\left( 
\begin{array}{cc}
1 & 1 \\ 
-1 & 1%
\end{array}
\right) \left( 
\begin{array}{cc}
1-\frac{S S^{\dagger}}{2} & S \\ 
-S^{\dagger} & 1-\frac{S^{\dagger} S}{2}%
\end{array}
\right)+O\left(S^3\right).
\end{eqnarray}
Under the conditions 
$
M_{D}^{T}=M_{D}, $ $
M_{N}S^{\dagger } =S^{T}M_{N} ,  $$
M_{N}S =S^{\ast }M_{N},  $$
M_{D}S^{\dagger } =S^{T}M_{D} ,  $$
M_{D}S =S^{\ast }M_{D},
$
we find 
\begin{eqnarray}
S &=&S^{\dag }=-\frac{1}{4}M_{D}^{-1}M_{N}, \\
S^{\ast } &=&S^{T}=-\frac{1}{4}M_{N}M_{D}^{-1}
\end{eqnarray}
and the mass matrices 
\begin{eqnarray}
U^{\ast }m^{\text{diag}}U^{\dagger } &=&
-M_{D}+\frac{M_{N}}{2}-\frac{1}{8}M_{N}M_{D}^{-1}M_{N},\\
V^{\ast }M^{\text{diag}}V^{\dagger }&=& M_{D}+\frac{M_{N}}{2}+\frac{1}{8}M_{N}M_{D}^{-1}M_{N}.%
\end{eqnarray}


\begin{thebibliography}{0}%
\makeatletter
\providecommand \@ifxundefined [1]{%
 \@ifx{#1\undefined}
}%
\providecommand \@ifnum [1]{%
 \ifnum #1\expandafter \@firstoftwo
 \else \expandafter \@secondoftwo
 \fi
}%
\providecommand \@ifx [1]{%
 \ifx #1\expandafter \@firstoftwo
 \else \expandafter \@secondoftwo
 \fi
}%
\providecommand \natexlab [1]{#1}%
\providecommand \enquote  [1]{``#1''}%
\providecommand \bibnamefont  [1]{#1}%
\providecommand \bibfnamefont [1]{#1}%
\providecommand \citenamefont [1]{#1}%
\providecommand \href@noop [0]{\@secondoftwo}%
\providecommand \href [0]{\begingroup \@sanitize@url \@href}%
\providecommand \@href[1]{\@@startlink{#1}\@@href}%
\providecommand \@@href[1]{\endgroup#1\@@endlink}%
\providecommand \@sanitize@url [0]{\catcode `\\12\catcode `\$12\catcode
  `\&12\catcode `\#12\catcode `\^12\catcode `\_12\catcode `\%12\relax}%
\providecommand \@@startlink[1]{}%
\providecommand \@@endlink[0]{}%
\providecommand \url  [0]{\begingroup\@sanitize@url \@url }%
\providecommand \@url [1]{\endgroup\@href {#1}{\urlprefix }}%
\providecommand \urlprefix  [0]{URL }%
\providecommand \Eprint [0]{\href }%
\@ifxundefined \urlstyle {%
  \providecommand \doi  [0]{\begingroup \@sanitize@url \@doi}%
  \providecommand \@doi [1]{\endgroup \@@startlink {\doibase
  #1}doi:\discretionary {}{}{}#1\@@endlink }%
}{%
  \providecommand \doi  [0]{doi:\discretionary{}{}{}\begingroup
  \urlstyle{rm}\Url }%
}%
\providecommand \doibase [0]{http://dx.doi.org/}%
\providecommand \Doi [0]{\begingroup \@sanitize@url \@Doi }%
\providecommand \@Doi  [1]{\endgroup\@@startlink{\doibase#1}\@@Doi}%
\providecommand \@@Doi [1]{#1\@@endlink}%
\providecommand \selectlanguage [0]{\@gobble}%
\providecommand \bibinfo  [0]{\@secondoftwo}%
\providecommand \bibfield  [0]{\@secondoftwo}%
\providecommand \translation [1]{[#1]}%
\providecommand \BibitemOpen [0]{}%
\providecommand \bibitemStop [0]{}%
\providecommand \bibitemNoStop [0]{.\EOS\space}%
\providecommand \EOS [0]{\spacefactor3000\relax}%
\providecommand \BibitemShut  [1]{\csname bibitem#1\endcsname}%
\end{thebibliography}%


\begin{thebibliography}{37}%
\makeatletter
\providecommand \@ifxundefined [1]{%
 \@ifx{#1\undefined}
}%
\providecommand \@ifnum [1]{%
 \ifnum #1\expandafter \@firstoftwo
 \else \expandafter \@secondoftwo
 \fi
}%
\providecommand \@ifx [1]{%
 \ifx #1\expandafter \@firstoftwo
 \else \expandafter \@secondoftwo
 \fi
}%
\providecommand \natexlab [1]{#1}%
\providecommand \enquote  [1]{``#1''}%
\providecommand \bibnamefont  [1]{#1}%
\providecommand \bibfnamefont [1]{#1}%
\providecommand \citenamefont [1]{#1}%
\providecommand \href@noop [0]{\@secondoftwo}%
\providecommand \href [0]{\begingroup \@sanitize@url \@href}%
\providecommand \@href[1]{\@@startlink{#1}\@@href}%
\providecommand \@@href[1]{\endgroup#1\@@endlink}%
\providecommand \@sanitize@url [0]{\catcode `\\12\catcode `\$12\catcode
  `\&12\catcode `\#12\catcode `\^12\catcode `\_12\catcode `\%12\relax}%
\providecommand \@@startlink[1]{}%
\providecommand \@@endlink[0]{}%
\providecommand \url  [0]{\begingroup\@sanitize@url \@url }%
\providecommand \@url [1]{\endgroup\@href {#1}{\urlprefix }}%
\providecommand \urlprefix  [0]{URL }%
\providecommand \Eprint [0]{\href }%
\providecommand \doibase [0]{http://dx.doi.org/}%
\providecommand \selectlanguage [0]{\@gobble}%
\providecommand \bibinfo  [0]{\@secondoftwo}%
\providecommand \bibfield  [0]{\@secondoftwo}%
\providecommand \translation [1]{[#1]}%
\providecommand \BibitemOpen [0]{}%
\providecommand \bibitemStop [0]{}%
\providecommand \bibitemNoStop [0]{.\EOS\space}%
\providecommand \EOS [0]{\spacefactor3000\relax}%
\providecommand \BibitemShut  [1]{\csname bibitem#1\endcsname}%
\let\auto@bib@innerbib\@empty
\bibitem [{\citenamefont {Nakamura}\ \emph {et~al.}(2010)\citenamefont
  {Nakamura} \emph {et~al.}}]{Nakamura:2010zzi}%
  \BibitemOpen
  \bibfield  {author} {\bibinfo {author} {\bibfnamefont {K.}~\bibnamefont
  {Nakamura}} \emph {et~al.} (\bibinfo {collaboration} {Particle Data Group}),\
  }\href {\doibase 10.1088/0954-3899/37/7A/075021} {\bibfield  {journal}
  {\bibinfo  {journal} {J.Phys.G}\ }\textbf {\bibinfo {volume} {G37}},\
  \bibinfo {pages} {075021} (\bibinfo {year} {2010})}\BibitemShut {NoStop}%
\bibitem [{\citenamefont {Cleveland}\ \emph {et~al.}(1998)\citenamefont
  {Cleveland}, \citenamefont {Daily}, \citenamefont {Davis}, \citenamefont
  {Distel}, \citenamefont {Lande} \emph {et~al.}}]{Cleveland:1998nv}%
  \BibitemOpen
  \bibfield  {author} {\bibinfo {author} {\bibfnamefont {B.}~\bibnamefont
  {Cleveland}}, \bibinfo {author} {\bibfnamefont {T.}~\bibnamefont {Daily}},
  \bibinfo {author} {\bibfnamefont {R.~J.}\ \bibnamefont {Davis}}, \bibinfo
  {author} {\bibfnamefont {J.~R.}\ \bibnamefont {Distel}}, \bibinfo {author}
  {\bibfnamefont {K.}~\bibnamefont {Lande}},  \emph {et~al.},\ }\href {\doibase
  10.1086/305343} {\bibfield  {journal} {\bibinfo  {journal} {Astrophys.J.}\
  }\textbf {\bibinfo {volume} {496}},\ \bibinfo {pages} {505} (\bibinfo {year}
  {1998})}\BibitemShut {NoStop}%
\bibitem [{\citenamefont {Fukuda}\ \emph {et~al.}(1996)\citenamefont {Fukuda}
  \emph {et~al.}}]{Fukuda:1996sz}%
  \BibitemOpen
  \bibfield  {author} {\bibinfo {author} {\bibfnamefont {Y.}~\bibnamefont
  {Fukuda}} \emph {et~al.} (\bibinfo {collaboration} {Kamiokande
  Collaboration}),\ }\href {\doibase 10.1103/PhysRevLett.77.1683} {\bibfield
  {journal} {\bibinfo  {journal} {Phys.Rev.Lett.}\ }\textbf {\bibinfo {volume}
  {77}},\ \bibinfo {pages} {1683} (\bibinfo {year} {1996})}\BibitemShut
  {NoStop}%
\bibitem [{\citenamefont {Abdurashitov}\ \emph {et~al.}(2009)\citenamefont
  {Abdurashitov} \emph {et~al.}}]{Abdurashitov:2009tn}%
  \BibitemOpen
  \bibfield  {author} {\bibinfo {author} {\bibfnamefont {J.}~\bibnamefont
  {Abdurashitov}} \emph {et~al.} (\bibinfo {collaboration} {SAGE
  Collaboration}),\ }\href {\doibase 10.1103/PhysRevC.80.015807} {\bibfield
  {journal} {\bibinfo  {journal} {Phys.Rev.}\ }\textbf {\bibinfo {volume}
  {C80}},\ \bibinfo {pages} {015807} (\bibinfo {year} {2009})},\ \Eprint
  {http://arxiv.org/abs/0901.2200} {arXiv:0901.2200 [nucl-ex]} \BibitemShut
  {NoStop}%
\bibitem [{\citenamefont {Anselmann}\ \emph {et~al.}(1992)\citenamefont
  {Anselmann} \emph {et~al.}}]{Anselmann:1992um}%
  \BibitemOpen
  \bibfield  {author} {\bibinfo {author} {\bibfnamefont {P.}~\bibnamefont
  {Anselmann}} \emph {et~al.} (\bibinfo {collaboration} {GALLEX
  Collaboration}),\ }\href {\doibase 10.1016/0370-2693(92)91521-A} {\bibfield
  {journal} {\bibinfo  {journal} {Phys.Lett.}\ }\textbf {\bibinfo {volume}
  {B285}},\ \bibinfo {pages} {376} (\bibinfo {year} {1992})}\BibitemShut
  {NoStop}%
\bibitem [{\citenamefont {Fukuda}\ \emph {et~al.}(2002)\citenamefont {Fukuda}
  \emph {et~al.}}]{Fukuda:2002pe}%
  \BibitemOpen
  \bibfield  {author} {\bibinfo {author} {\bibfnamefont {S.}~\bibnamefont
  {Fukuda}} \emph {et~al.} (\bibinfo {collaboration} {Super-Kamiokande
  Collaboration}),\ }\href {\doibase 10.1016/S0370-2693(02)02090-7} {\bibfield
  {journal} {\bibinfo  {journal} {Phys.Lett.}\ }\textbf {\bibinfo {volume}
  {B539}},\ \bibinfo {pages} {179} (\bibinfo {year} {2002})},\ \Eprint
  {http://arxiv.org/abs/hep-ex/0205075} {arXiv:hep-ex/0205075 [hep-ex]}
  \BibitemShut {NoStop}%
\bibitem [{\citenamefont {Ahmad}\ \emph {et~al.}(2001)\citenamefont {Ahmad}
  \emph {et~al.}}]{Ahmad:2001an}%
  \BibitemOpen
  \bibfield  {author} {\bibinfo {author} {\bibfnamefont {Q.}~\bibnamefont
  {Ahmad}} \emph {et~al.} (\bibinfo {collaboration} {SNO Collaboration}),\
  }\href {\doibase 10.1103/PhysRevLett.87.071301} {\bibfield  {journal}
  {\bibinfo  {journal} {Phys.Rev.Lett.}\ }\textbf {\bibinfo {volume} {87}},\
  \bibinfo {pages} {071301} (\bibinfo {year} {2001})},\ \Eprint
  {http://arxiv.org/abs/nucl-ex/0106015} {arXiv:nucl-ex/0106015 [nucl-ex]}
  \BibitemShut {NoStop}%
\bibitem [{\citenamefont {Fukuda}\ \emph {et~al.}(1998)\citenamefont {Fukuda}
  \emph {et~al.}}]{Fukuda:1998mi}%
  \BibitemOpen
  \bibfield  {author} {\bibinfo {author} {\bibfnamefont {Y.}~\bibnamefont
  {Fukuda}} \emph {et~al.} (\bibinfo {collaboration} {Super-Kamiokande
  Collaboration}),\ }\href {\doibase 10.1103/PhysRevLett.81.1562} {\bibfield
  {journal} {\bibinfo  {journal} {Phys.Rev.Lett.}\ }\textbf {\bibinfo {volume}
  {81}},\ \bibinfo {pages} {1562} (\bibinfo {year} {1998})},\ \Eprint
  {http://arxiv.org/abs/hep-ex/9807003} {arXiv:hep-ex/9807003 [hep-ex]}
  \BibitemShut {NoStop}%
\bibitem [{\citenamefont {Ashie}\ \emph {et~al.}(2004)\citenamefont {Ashie}
  \emph {et~al.}}]{Ashie:2004mr}%
  \BibitemOpen
  \bibfield  {author} {\bibinfo {author} {\bibfnamefont {Y.}~\bibnamefont
  {Ashie}} \emph {et~al.} (\bibinfo {collaboration} {Super-Kamiokande
  Collaboration}),\ }\href {\doibase 10.1103/PhysRevLett.93.101801} {\bibfield
  {journal} {\bibinfo  {journal} {Phys.Rev.Lett.}\ }\textbf {\bibinfo {volume}
  {93}},\ \bibinfo {pages} {101801} (\bibinfo {year} {2004})},\ \Eprint
  {http://arxiv.org/abs/hep-ex/0404034} {arXiv:hep-ex/0404034 [hep-ex]}
  \BibitemShut {NoStop}%
\bibitem [{\citenamefont {Eguchi}\ \emph {et~al.}(2003)\citenamefont {Eguchi}
  \emph {et~al.}}]{Eguchi:2002dm}%
  \BibitemOpen
  \bibfield  {author} {\bibinfo {author} {\bibfnamefont {K.}~\bibnamefont
  {Eguchi}} \emph {et~al.} (\bibinfo {collaboration} {KamLAND Collaboration}),\
  }\href {\doibase 10.1103/PhysRevLett.90.021802} {\bibfield  {journal}
  {\bibinfo  {journal} {Phys.Rev.Lett.}\ }\textbf {\bibinfo {volume} {90}},\
  \bibinfo {pages} {021802} (\bibinfo {year} {2003})},\ \Eprint
  {http://arxiv.org/abs/hep-ex/0212021} {arXiv:hep-ex/0212021 [hep-ex]}
  \BibitemShut {NoStop}%
\bibitem [{\citenamefont {Araki}\ \emph {et~al.}(2005)\citenamefont {Araki}
  \emph {et~al.}}]{Araki:2004mb}%
  \BibitemOpen
  \bibfield  {author} {\bibinfo {author} {\bibfnamefont {T.}~\bibnamefont
  {Araki}} \emph {et~al.} (\bibinfo {collaboration} {KamLAND Collaboration}),\
  }\href {\doibase 10.1103/PhysRevLett.94.081801} {\bibfield  {journal}
  {\bibinfo  {journal} {Phys.Rev.Lett.}\ }\textbf {\bibinfo {volume} {94}},\
  \bibinfo {pages} {081801} (\bibinfo {year} {2005})},\ \Eprint
  {http://arxiv.org/abs/hep-ex/0406035} {arXiv:hep-ex/0406035 [hep-ex]}
  \BibitemShut {NoStop}%
\bibitem [{\citenamefont {Arpesella}\ \emph {et~al.}(2008)\citenamefont
  {Arpesella} \emph {et~al.}}]{Arpesella:2008mt}%
  \BibitemOpen
  \bibfield  {author} {\bibinfo {author} {\bibfnamefont {C.}~\bibnamefont
  {Arpesella}} \emph {et~al.} (\bibinfo {collaboration} {The Borexino
  Collaboration}),\ }\href {\doibase 10.1103/PhysRevLett.101.091302} {\bibfield
   {journal} {\bibinfo  {journal} {Phys.Rev.Lett.}\ }\textbf {\bibinfo {volume}
  {101}},\ \bibinfo {pages} {091302} (\bibinfo {year} {2008})},\ \Eprint
  {http://arxiv.org/abs/0805.3843} {arXiv:0805.3843 [astro-ph]} \BibitemShut
  {NoStop}%
\bibitem [{\citenamefont {Ahn}\ \emph {et~al.}(2006)\citenamefont {Ahn} \emph
  {et~al.}}]{Ahn:2006zza}%
  \BibitemOpen
  \bibfield  {author} {\bibinfo {author} {\bibfnamefont {M.}~\bibnamefont
  {Ahn}} \emph {et~al.} (\bibinfo {collaboration} {K2K Collaboration}),\ }\href
  {\doibase 10.1103/PhysRevD.74.072003} {\bibfield  {journal} {\bibinfo
  {journal} {Phys.Rev.}\ }\textbf {\bibinfo {volume} {D74}},\ \bibinfo {pages}
  {072003} (\bibinfo {year} {2006})},\ \Eprint
  {http://arxiv.org/abs/hep-ex/0606032} {arXiv:hep-ex/0606032 [hep-ex]}
  \BibitemShut {NoStop}%
\bibitem [{\citenamefont {Michael}\ \emph {et~al.}(2006)\citenamefont {Michael}
  \emph {et~al.}}]{Michael:2006rx}%
  \BibitemOpen
  \bibfield  {author} {\bibinfo {author} {\bibfnamefont {D.}~\bibnamefont
  {Michael}} \emph {et~al.} (\bibinfo {collaboration} {MINOS Collaboration}),\
  }\href {\doibase 10.1103/PhysRevLett.97.191801} {\bibfield  {journal}
  {\bibinfo  {journal} {Phys.Rev.Lett.}\ }\textbf {\bibinfo {volume} {97}},\
  \bibinfo {pages} {191801} (\bibinfo {year} {2006})},\ \Eprint
  {http://arxiv.org/abs/hep-ex/0607088} {arXiv:hep-ex/0607088 [hep-ex]}
  \BibitemShut {NoStop}%
\bibitem [{\citenamefont {Cortez}\ and\ \citenamefont
  {Tonasse}(2005)}]{Cortez:2005cp}%
  \BibitemOpen
  \bibfield  {author} {\bibinfo {author} {\bibfnamefont {N.~V.}\ \bibnamefont
  {Cortez}}\ and\ \bibinfo {author} {\bibfnamefont {M.~D.}\ \bibnamefont
  {Tonasse}},\ }\href {\doibase 10.1103/PhysRevD.72.073005} {\bibfield
  {journal} {\bibinfo  {journal} {Phys.Rev.}\ }\textbf {\bibinfo {volume}
  {D72}},\ \bibinfo {pages} {073005} (\bibinfo {year} {2005})},\ \Eprint
  {http://arxiv.org/abs/hep-ph/0510143} {arXiv:hep-ph/0510143 [hep-ph]}
  \BibitemShut {NoStop}%
\bibitem [{\citenamefont {Dias}\ \emph {et~al.}(2011)\citenamefont {Dias},
  \citenamefont {{de S. Pires}},\ and\ \citenamefont {{da
  Silva}}}]{Dias:2011sq}%
  \BibitemOpen
  \bibfield  {author} {\bibinfo {author} {\bibfnamefont {A.}~\bibnamefont
  {Dias}}, \bibinfo {author} {\bibfnamefont {C.}~\bibnamefont {{de S. Pires}}},
  \ and\ \bibinfo {author} {\bibfnamefont {P.~R.}\ \bibnamefont {{da Silva}}},\
  }\href {\doibase 10.1103/PhysRevD.84.053011} {\bibfield  {journal} {\bibinfo
  {journal} {Phys.Rev.}\ }\textbf {\bibinfo {volume} {D84}},\ \bibinfo {pages}
  {053011} (\bibinfo {year} {2011})},\ \Eprint {http://arxiv.org/abs/1107.0739}
  {arXiv:1107.0739 [hep-ph]} \BibitemShut {NoStop}%
\bibitem [{\citenamefont {Dong}\ and\ \citenamefont
  {Long}(2008)}]{Dong:2008sw}%
  \BibitemOpen
  \bibfield  {author} {\bibinfo {author} {\bibfnamefont {P.}~\bibnamefont
  {Dong}}\ and\ \bibinfo {author} {\bibfnamefont {H.~N.}\ \bibnamefont
  {Long}},\ }\href {\doibase 10.1103/PhysRevD.77.057302} {\bibfield  {journal}
  {\bibinfo  {journal} {Phys.Rev.}\ }\textbf {\bibinfo {volume} {D77}},\
  \bibinfo {pages} {057302} (\bibinfo {year} {2008})},\ \Eprint
  {http://arxiv.org/abs/0801.4196} {arXiv:0801.4196 [hep-ph]} \BibitemShut
  {NoStop}%
\bibitem [{\citenamefont {Ibarra}\ \emph {et~al.}(2010)\citenamefont {Ibarra},
  \citenamefont {Molinaro},\ and\ \citenamefont {Petcov}}]{Ibarra:2010xw}%
  \BibitemOpen
  \bibfield  {author} {\bibinfo {author} {\bibfnamefont {A.}~\bibnamefont
  {Ibarra}}, \bibinfo {author} {\bibfnamefont {E.}~\bibnamefont {Molinaro}}, \
  and\ \bibinfo {author} {\bibfnamefont {S.}~\bibnamefont {Petcov}},\ }\href
  {\doibase 10.1007/JHEP09(2010)108} {\bibfield  {journal} {\bibinfo  {journal}
  {JHEP}\ }\textbf {\bibinfo {volume} {1009}},\ \bibinfo {pages} {108}
  (\bibinfo {year} {2010})},\ \Eprint {http://arxiv.org/abs/1007.2378}
  {arXiv:1007.2378 [hep-ph]} \BibitemShut {NoStop}%
\bibitem [{\citenamefont {Gu}\ and\ \citenamefont {Sarkar}(2010)}]{Gu:2010xc}%
  \BibitemOpen
  \bibfield  {author} {\bibinfo {author} {\bibfnamefont {P.-H.}\ \bibnamefont
  {Gu}}\ and\ \bibinfo {author} {\bibfnamefont {U.}~\bibnamefont {Sarkar}},\
  }\href {\doibase 10.1016/j.physletb.2010.09.062} {\bibfield  {journal}
  {\bibinfo  {journal} {Phys.Lett.}\ }\textbf {\bibinfo {volume} {B694}},\
  \bibinfo {pages} {226} (\bibinfo {year} {2010})},\ \Eprint
  {http://arxiv.org/abs/1007.2323} {arXiv:1007.2323 [hep-ph]} \BibitemShut
  {NoStop}%
\bibitem [{\citenamefont {Diaz}\ \emph {et~al.}(2004)\citenamefont {Diaz},
  \citenamefont {Martinez},\ and\ \citenamefont {Ochoa}}]{Diaz:2003dk}%
  \BibitemOpen
  \bibfield  {author} {\bibinfo {author} {\bibfnamefont {R.~A.}\ \bibnamefont
  {Diaz}}, \bibinfo {author} {\bibfnamefont {R.}~\bibnamefont {Martinez}}, \
  and\ \bibinfo {author} {\bibfnamefont {F.}~\bibnamefont {Ochoa}},\ }\href
  {\doibase 10.1103/PhysRevD.69.095009} {\bibfield  {journal} {\bibinfo
  {journal} {Phys.Rev.}\ }\textbf {\bibinfo {volume} {D69}},\ \bibinfo {pages}
  {095009} (\bibinfo {year} {2004})},\ \Eprint
  {http://arxiv.org/abs/hep-ph/0309280} {arXiv:hep-ph/0309280 [hep-ph]}
  \BibitemShut {NoStop}%
\bibitem [{\citenamefont {Diaz}\ \emph {et~al.}(2005)\citenamefont {Diaz},
  \citenamefont {Martinez},\ and\ \citenamefont {Ochoa}}]{Diaz:2004fs}%
  \BibitemOpen
  \bibfield  {author} {\bibinfo {author} {\bibfnamefont {R.~A.}\ \bibnamefont
  {Diaz}}, \bibinfo {author} {\bibfnamefont {R.}~\bibnamefont {Martinez}}, \
  and\ \bibinfo {author} {\bibfnamefont {F.}~\bibnamefont {Ochoa}},\ }\href
  {\doibase 10.1103/PhysRevD.72.035018} {\bibfield  {journal} {\bibinfo
  {journal} {Phys.Rev.}\ }\textbf {\bibinfo {volume} {D72}},\ \bibinfo {pages}
  {035018} (\bibinfo {year} {2005})},\ \Eprint
  {http://arxiv.org/abs/hep-ph/0411263} {arXiv:hep-ph/0411263 [hep-ph]}
  \BibitemShut {NoStop}%
\bibitem [{\citenamefont {Duong}\ and\ \citenamefont
  {Ma}(1993)}]{Duong:1993zn}%
  \BibitemOpen
  \bibfield  {author} {\bibinfo {author} {\bibfnamefont {T.}~\bibnamefont
  {Duong}}\ and\ \bibinfo {author} {\bibfnamefont {E.}~\bibnamefont {Ma}},\
  }\href {\doibase 10.1016/0370-2693(93)90329-G} {\bibfield  {journal}
  {\bibinfo  {journal} {Phys.Lett.}\ }\textbf {\bibinfo {volume} {B316}},\
  \bibinfo {pages} {307} (\bibinfo {year} {1993})},\ \Eprint
  {http://arxiv.org/abs/hep-ph/9306264} {arXiv:hep-ph/9306264 [hep-ph]}
  \BibitemShut {NoStop}%
\bibitem [{\citenamefont {Montero}\ \emph {et~al.}(2002)\citenamefont
  {Montero}, \citenamefont {de~S.~Pires},\ and\ \citenamefont
  {Pleitez}}]{Montero:2001ji}%
  \BibitemOpen
  \bibfield  {author} {\bibinfo {author} {\bibfnamefont {J.}~\bibnamefont
  {Montero}}, \bibinfo {author} {\bibfnamefont {C.}~\bibnamefont
  {de~S.~Pires}}, \ and\ \bibinfo {author} {\bibfnamefont {V.}~\bibnamefont
  {Pleitez}},\ }\href {\doibase 10.1103/PhysRevD.65.093017} {\bibfield
  {journal} {\bibinfo  {journal} {Phys.Rev.}\ }\textbf {\bibinfo {volume}
  {D65}},\ \bibinfo {pages} {093017} (\bibinfo {year} {2002})},\ \Eprint
  {http://arxiv.org/abs/hep-ph/0103096} {arXiv:hep-ph/0103096 [hep-ph]}
  \BibitemShut {NoStop}%
\bibitem [{\citenamefont {Tully}\ and\ \citenamefont
  {Joshi}(2001)}]{Tully:2000kk}%
  \BibitemOpen
  \bibfield  {author} {\bibinfo {author} {\bibfnamefont {M.}~\bibnamefont
  {Tully}}\ and\ \bibinfo {author} {\bibfnamefont {G.~C.}\ \bibnamefont
  {Joshi}},\ }\href {\doibase 10.1103/PhysRevD.64.011301} {\bibfield  {journal}
  {\bibinfo  {journal} {Phys.Rev.}\ }\textbf {\bibinfo {volume} {D64}},\
  \bibinfo {pages} {011301} (\bibinfo {year} {2001})},\ \Eprint
  {http://arxiv.org/abs/hep-ph/0011172} {arXiv:hep-ph/0011172 [hep-ph]}
  \BibitemShut {NoStop}%
\bibitem [{\citenamefont {Dong}\ \emph {et~al.}(2010)\citenamefont {Dong},
  \citenamefont {Hue}, \citenamefont {Long},\ and\ \citenamefont
  {Soa}}]{Dong:2010gk}%
  \BibitemOpen
  \bibfield  {author} {\bibinfo {author} {\bibfnamefont {P.}~\bibnamefont
  {Dong}}, \bibinfo {author} {\bibfnamefont {L.}~\bibnamefont {Hue}}, \bibinfo
  {author} {\bibfnamefont {H.}~\bibnamefont {Long}}, \ and\ \bibinfo {author}
  {\bibfnamefont {D.}~\bibnamefont {Soa}},\ }\href {\doibase
  10.1103/PhysRevD.81.053004} {\bibfield  {journal} {\bibinfo  {journal}
  {Phys.Rev.}\ }\textbf {\bibinfo {volume} {D81}},\ \bibinfo {pages} {053004}
  (\bibinfo {year} {2010})},\ \Eprint {http://arxiv.org/abs/1001.4625}
  {arXiv:1001.4625 [hep-ph]} \BibitemShut {NoStop}%
\bibitem [{\citenamefont {Yin}(2007)}]{Yin:2007rv}%
  \BibitemOpen
  \bibfield  {author} {\bibinfo {author} {\bibfnamefont {F.}~\bibnamefont
  {Yin}},\ }\href {\doibase 10.1103/PhysRevD.75.073010} {\bibfield  {journal}
  {\bibinfo  {journal} {Phys.Rev.}\ }\textbf {\bibinfo {volume} {D75}},\
  \bibinfo {pages} {073010} (\bibinfo {year} {2007})},\ \Eprint
  {http://arxiv.org/abs/0704.3827} {arXiv:0704.3827 [hep-ph]} \BibitemShut
  {NoStop}%
\bibitem [{\citenamefont {Dias}\ \emph {et~al.}(2005)\citenamefont {Dias},
  \citenamefont {de~S.~Pires},\ and\ \citenamefont {Rodrigues~da
  Silva}}]{Dias:2005yh}%
  \BibitemOpen
  \bibfield  {author} {\bibinfo {author} {\bibfnamefont {A.~G.}\ \bibnamefont
  {Dias}}, \bibinfo {author} {\bibfnamefont {C.}~\bibnamefont {de~S.~Pires}}, \
  and\ \bibinfo {author} {\bibfnamefont {P.}~\bibnamefont {Rodrigues~da
  Silva}},\ }\href {\doibase 10.1016/j.physletb.2005.09.028} {\bibfield
  {journal} {\bibinfo  {journal} {Phys.Lett.}\ }\textbf {\bibinfo {volume}
  {B628}},\ \bibinfo {pages} {85} (\bibinfo {year} {2005})},\ \Eprint
  {http://arxiv.org/abs/hep-ph/0508186} {arXiv:hep-ph/0508186 [hep-ph]}
  \BibitemShut {NoStop}%
\bibitem [{\citenamefont {Cogollo}\ \emph {et~al.}(2008)\citenamefont
  {Cogollo}, \citenamefont {Diniz}, \citenamefont {de~S.~Pires},\ and\
  \citenamefont {Rodrigues~da Silva}}]{Cogollo:2008zc}%
  \BibitemOpen
  \bibfield  {author} {\bibinfo {author} {\bibfnamefont {D.}~\bibnamefont
  {Cogollo}}, \bibinfo {author} {\bibfnamefont {H.}~\bibnamefont {Diniz}},
  \bibinfo {author} {\bibfnamefont {C.}~\bibnamefont {de~S.~Pires}}, \ and\
  \bibinfo {author} {\bibfnamefont {P.}~\bibnamefont {Rodrigues~da Silva}},\
  }\href {\doibase 10.1140/epjc/s10052-008-0749-5} {\bibfield  {journal}
  {\bibinfo  {journal} {Eur.Phys.J.}\ }\textbf {\bibinfo {volume} {C58}},\
  \bibinfo {pages} {455} (\bibinfo {year} {2008})},\ \Eprint
  {http://arxiv.org/abs/0806.3087} {arXiv:0806.3087 [hep-ph]} \BibitemShut
  {NoStop}%
\bibitem [{\citenamefont {Cogollo}\ \emph {et~al.}(2009)\citenamefont
  {Cogollo}, \citenamefont {Diniz},\ and\ \citenamefont
  {de~S.~Pires}}]{Cogollo:2009yi}%
  \BibitemOpen
  \bibfield  {author} {\bibinfo {author} {\bibfnamefont {D.}~\bibnamefont
  {Cogollo}}, \bibinfo {author} {\bibfnamefont {H.}~\bibnamefont {Diniz}}, \
  and\ \bibinfo {author} {\bibfnamefont {C.}~\bibnamefont {de~S.~Pires}},\
  }\href@noop {} {\enquote {\bibinfo {title} {{KeV right-handed neutrinos from
  type II seesaw mechanism in a 3-3-1 model}},}\ } (\bibinfo {year} {2009}),\
  \Eprint {http://arxiv.org/abs/0903.0370} {arXiv:0903.0370 [hep-ph]}
  \BibitemShut {NoStop}%
\bibitem [{\citenamefont {Martinez}\ and\ \citenamefont
  {Ochoa}(2007)}]{Martinez:2006gb}%
  \BibitemOpen
  \bibfield  {author} {\bibinfo {author} {\bibfnamefont {R.}~\bibnamefont
  {Martinez}}\ and\ \bibinfo {author} {\bibfnamefont {F.}~\bibnamefont
  {Ochoa}},\ }\href {\doibase 10.1140/epjc/s10052-007-0307-6} {\bibfield
  {journal} {\bibinfo  {journal} {Eur.Phys.J.}\ }\textbf {\bibinfo {volume}
  {C51}},\ \bibinfo {pages} {701} (\bibinfo {year} {2007})},\ \Eprint
  {http://arxiv.org/abs/hep-ph/0606173} {arXiv:hep-ph/0606173 [hep-ph]}
  \BibitemShut {NoStop}%
\bibitem [{\citenamefont {Barr}(2004)}]{Barr:2003nn}%
  \BibitemOpen
  \bibfield  {author} {\bibinfo {author} {\bibfnamefont {S.}~\bibnamefont
  {Barr}},\ }\href {\doibase 10.1103/PhysRevLett.92.101601} {\bibfield
  {journal} {\bibinfo  {journal} {Phys.Rev.Lett.}\ }\textbf {\bibinfo {volume}
  {92}},\ \bibinfo {pages} {101601} (\bibinfo {year} {2004})},\ \Eprint
  {http://arxiv.org/abs/hep-ph/0309152} {arXiv:hep-ph/0309152 [hep-ph]}
  \BibitemShut {NoStop}%
\bibitem [{\citenamefont {Mohapatra}(1986)}]{Mohapatra:1986aw}%
  \BibitemOpen
  \bibfield  {author} {\bibinfo {author} {\bibfnamefont {R.}~\bibnamefont
  {Mohapatra}},\ }\href {\doibase 10.1103/PhysRevLett.56.561} {\bibfield
  {journal} {\bibinfo  {journal} {Phys.Rev.Lett.}\ }\textbf {\bibinfo {volume}
  {56}},\ \bibinfo {pages} {561} (\bibinfo {year} {1986})}\BibitemShut
  {NoStop}%
\bibitem [{\citenamefont {Mohapatra}\ and\ \citenamefont
  {Valle}(1986)}]{Mohapatra:1986bd}%
  \BibitemOpen
  \bibfield  {author} {\bibinfo {author} {\bibfnamefont {R.}~\bibnamefont
  {Mohapatra}}\ and\ \bibinfo {author} {\bibfnamefont {J.}~\bibnamefont
  {Valle}},\ }\href {\doibase 10.1103/PhysRevD.34.1642} {\bibfield  {journal}
  {\bibinfo  {journal} {Phys.Rev.}\ }\textbf {\bibinfo {volume} {D34}},\
  \bibinfo {pages} {1642} (\bibinfo {year} {1986})}\BibitemShut {NoStop}%
\bibitem [{\citenamefont {King}(2002)}]{King:2002nf}%
  \BibitemOpen
  \bibfield  {author} {\bibinfo {author} {\bibfnamefont {S.}~\bibnamefont
  {King}},\ }\href@noop {} {\bibfield  {journal} {\bibinfo  {journal} {JHEP}\
  }\textbf {\bibinfo {volume} {0209}},\ \bibinfo {pages} {011} (\bibinfo {year}
  {2002})},\ \Eprint {http://arxiv.org/abs/hep-ph/0204360}
  {arXiv:hep-ph/0204360 [hep-ph]} \BibitemShut {NoStop}%
\bibitem [{\citenamefont {Strumia}\ and\ \citenamefont
  {Vissani}(2006)}]{Strumia:2006db}%
  \BibitemOpen
  \bibfield  {author} {\bibinfo {author} {\bibfnamefont {A.}~\bibnamefont
  {Strumia}}\ and\ \bibinfo {author} {\bibfnamefont {F.}~\bibnamefont
  {Vissani}},\ }\href@noop {} {\  (\bibinfo {year} {2006})},\ \Eprint
  {http://arxiv.org/abs/hep-ph/0606054} {arXiv:hep-ph/0606054 [hep-ph]}
  \BibitemShut {NoStop}%
\bibitem [{\citenamefont {Grimus}\ \emph {et~al.}(2004)\citenamefont {Grimus},
  \citenamefont {Joshipura}, \citenamefont {Lavoura},\ and\ \citenamefont
  {Tanimoto}}]{Grimus:2004hf}%
  \BibitemOpen
  \bibfield  {author} {\bibinfo {author} {\bibfnamefont {W.}~\bibnamefont
  {Grimus}}, \bibinfo {author} {\bibfnamefont {A.~S.}\ \bibnamefont
  {Joshipura}}, \bibinfo {author} {\bibfnamefont {L.}~\bibnamefont {Lavoura}},
  \ and\ \bibinfo {author} {\bibfnamefont {M.}~\bibnamefont {Tanimoto}},\
  }\href {\doibase 10.1140/epjc/s2004-01890-5} {\bibfield  {journal} {\bibinfo
  {journal} {Eur.Phys.J.}\ }\textbf {\bibinfo {volume} {C36}},\ \bibinfo
  {pages} {227} (\bibinfo {year} {2004})},\ \Eprint
  {http://arxiv.org/abs/hep-ph/0405016} {arXiv:hep-ph/0405016 [hep-ph]}
  \BibitemShut {NoStop}%
\bibitem [{\citenamefont {Altarelli}(2007)}]{Altarelli:2007gb}%
  \BibitemOpen
  \bibfield  {author} {\bibinfo {author} {\bibfnamefont {G.}~\bibnamefont
  {Altarelli}},\ }\href@noop {} {\  (\bibinfo {year} {2007})},\ \Eprint
  {http://arxiv.org/abs/0711.0161} {arXiv:0711.0161 [hep-ph]} \BibitemShut
  {NoStop}%
\end{thebibliography}
\end{document}